\documentclass[anonymous=false]{acmart}

\usepackage[utf8]{inputenc}
\usepackage{multirow}
\usepackage{xcolor,colortbl} 
\usepackage{wrapfig}
\usepackage{markdown}
\usepackage{subcaption}
\usepackage[textwidth=1.5cm]{todonotes}
\usepackage{natbib}
\usepackage{graphicx} 

\copyrightyear{2023}
\setcopyright{none}

\begin{document}

\title{Exploring Social Choice Mechanisms for Recommendation Fairness in SCRUF}

\author[A. Aird]{Amanda Aird}
\email{amanda.aird@colorado.edu}
\affiliation{%
  \institution{Department of Information Science; University of Colorado, Boulder}
  \streetaddress{}
  \city{Boulder}
  \state{Colorado}
  \country{USA}
  \postcode{80309}
}

\author[C. All]{Cassidy All}
\email{cassidy.all@colorado.edu}
\affiliation{%
  \institution{Department of Information Science; University of Colorado, Boulder}
  \streetaddress{}
  \city{Boulder}
  \state{Colorado}
  \country{USA}
  \postcode{80309}
}

\author[P. Farastu]{Paresha Farastu}
\email{paresha.farastu@colorado.edu}
\affiliation{%
  \institution{Department of Computer Science; University of Colorado, Boulder}
  \streetaddress{}
  \city{Boulder}
  \state{Colorado}
  \country{USA}
  \postcode{80309}
}

\author[E. \u{S}tefancov\'{a}]{Elena \u{S}tefancov\'{a}}
\email{elena.stefancova@fmph.uniba.sk}
\affiliation{%
  \institution{Comenius University Bratislava}
  \streetaddress{}
  \city{Bratislava}
  \country{Slovakia}
}

\author[J. Sun]{Joshua Sun}
\email{joshua.sun@colorado.edu}
\affiliation{%
  \institution{Independent Researcher}
  \streetaddress{}
  \city{Boulder}
  \state{Colorado}
  \country{USA}
  \postcode{80309}
}

\author[N. Mattei]{Nicholas Mattei}
\email{nsmattei@tulane.edu}
\affiliation{%
  \institution{Department of Computer Science; Tulane University}
  \streetaddress{}
  \city{New Orleans}
  \state{Louisiana}
  \country{USA}
  \postcode{70118}
}

\author[R. Burke]{Robin Burke}
\email{robin.burke@colorado.edu}
\orcid{0000-0001-5766-6434}
\affiliation{%
  \institution{Department of Information Science; University of Colorado, Boulder}
  \streetaddress{}
  \city{Boulder}
  \state{Colorado}
  \country{USA}
  \postcode{80309}
}

\begin{abstract}
Fairness problems in recommender systems often have a complexity in practice that is not adequately captured in simplified research formulations. A social choice formulation of the fairness problem, operating within a multi-agent architecture of fairness concerns, offers a flexible and multi-aspect alternative to fairness-aware recommendation approaches. Leveraging social choice allows for increased generality and the possibility of tapping into well-studied social choice algorithms for resolving the tension between multiple, competing fairness concerns. This paper explores a range of options for choice mechanisms in multi-aspect fairness applications using both real and synthetic data and shows that different classes of choice and allocation mechanisms yield different but consistent fairness / accuracy tradeoffs. We also show that a multi-agent formulation offers flexibility in adapting to user population dynamics.
\end{abstract}

\maketitle

\section{Introduction}
The complexity of fairness considerations in recommender systems is well understood; see \citet{ekstrand2022fairness} for a survey. Practical applications involving fairness require attention to multiple fairness concerns, each potentially formulated in a different way, relevant to a different set of stakeholders \cite{smith2023many}.  Methods that assume a single dimension of fairness or that assume all fairness concerns are formulated identically will not be successful in these applications.

The SCRUF-D architecture outlined in \citet{aird2023dynamic} and \citet{burke2022multi}
offers one possible way to address the complexity of fairness-aware recommendation by formulating it as a two-phase social choice problem within an architecture where fairness concerns are represented as agents. A fairness concern calls out one or more features and designates a set of protected values for these features. Each of these fairness concerns also articulates a function that takes a recommendation history $L_{t-1}$ and generates a value in $[0,1]$, where 0 is maximally unfair and 1 is the fairness target.

Each of these fairness concerns can be instantiated as an agent that can take as input the user preference ranking along with $L$ and produce a ranking over the set of items. In the first phase, fairness agents are allocated to recommendation opportunities, i.e., user arrivals. Performing this online allocation of fairness agents allows the system to adapt dynamically to fairness outcomes as users are served recommendations. In the second phase, allocated agents and the core recommendation algorithm contribute preferences over items to a preference aggregation mechanism, i.e., a voting rule, the output of which is a ranked list of recommended items to be delivered to the user.


The key advantage of the SCRUF-D approach is its generality. Fairness agents can be implemented in many different ways, with different objective functions and measures, while both the allocation and aggregation mechanisms can be chosen from the large variety of mechanisms that have been studied in computational social choice and for which formal properties are known \cite{BCELP16a}. \footnote{Note that not all properties of interest to social choice theorists are relevant to the fairness-aware recommendation scenario. For example, in rivalrous settings, an algorithm's potential for preference revelation might be of concern. In the SCRUF setting, agents are assumed to be all developed within a single organization for the goal of fair recommendation and are not considered rivalrous.}
In this paper, we explore some of the many choices a system implementer has for these mechanisms. For the allocation phase, we examine lottery mechanisms (in which a single agent is chosen from a dynamically generated lottery distribution), weighted mechanisms (in which all agents are allocated with dynamically computed weights), and a simple least misery allocation. For preference aggregation, we study two weighted/score based methods (weighted voting and Borda score) and two pair-wise methods (Copeland and RankedPairs).

In particular, we ask the following research questions: \\
\textbf{RQ 1}: Do different mechanisms offer different fairness / accuracy tradeoffs for different conditions, and if so, why?\\
\textbf{RQ 2}: Do different social choice mechanisms have different dynamic characteristics, and if so, why? \\
 \textbf{RQ 3}: What is the interaction between the mechanisms for allocation and choice and what makes for good synergy?

\section{Related Work}

Within the field of (computational) social choice there have been several investigations into the idea of \emph{dynamic} settings, which resemble the problem one faces in a recommender system where users arrive online. \citet{freeman2017fair} investigate what they call \emph{dynamic social choice functions} in settings where a fixed set of agents select a single item to share over a series of time steps. \citet{lackner2020perpetual} study the problem of voting (selecting a single outcome to share) over multiple time steps where various properties such as fairness need to be guaranteed over the total time horizon.   \citet{parkes2013dynamic} look at social choice settings where the preferences of agents evolve over time in response to the outcome of past rounds of voting. However, in all these settings the whole set of agents shares the resulting decision, whereas we are focused on fairness over sets of individual personalized recommendations.

The architecture presented here advances and generalizes the approach found in \cite{sonboli2020and}. Like that architecture, fairness concerns are represented as agents and interact through social choice. However, in \cite{sonboli2020and}, the allocation mechanism selects only a single agent at each time step and the choice mechanism has a fixed, additive, form. We allow for a wider variety of allocation and choice mechanisms, and therefore present a more general solution.

\citet{ge2021towards} investigate the problem of long term dynamic fairness in recommendation systems. This work, like ours, highlights the need to ensure that fairness is preserved as a temporal concept 
To this end they propose a framework to ensure fairness of exposure to the producers of items by casting the problem as a constrained Markov Decision Process where the actions are recommendations and the reward function takes into account both utility and exposure. As above, this work fixes definitions of fairness a priori, although their learning methodology may serve as inspiration to future extensions of our work.

\citet{morik2020controlling} investigate the problem of learning to rank over large item sets while ensuring fairness of merit based guarantees to groups of item producers. Specifically, they adapt existing methods to ensure that the exposure is \emph{unbiased}, e.g., not subject to rich-get-richer dynamics, and \emph{fairness} defined as exposure being proportional to merit. Both of these goals are built into the regularization of the learner. In contrast, our work factors out the recommendation methodology and we encapsulate fairness definitions as separate agents rather than embedded in the learning objective, allowing our framework to be more flexible.

This paper concentrates on provider-side fairness \cite{burke_multisided_2017} although SCRUF-D is intended to be compatible with multisided fairness as well. There have been a number of efforts that explicitly consider the multisided nature of fairness in recommendation and matching platforms. \citet{patro2020fairrec} investigate fairness in two-sided matching platforms where there are both producers and consumers with different definitions of fairness. \citet{patro2020fairrec} also appeal to the literature on the fair allocation of indivisible goods from the social choice literature \cite{Thomson:FairRules}. Their work is closest to the allocation phase of our algorithm. However, in contrast to our work they only use exposure on the producer side and relevance on the consumer side as fairness metrics, whereas our work aims to capture additional definitions. 

An important distinction between the work we present here and that of \citet{patro2020fairrec} is that our algorithms operate online, as users arrive, whereas \citet{patro2020fairrec} use a batch technique to create and cache a fair distribution of the whole catalog of users and items. Batch techniques cannot guarantee that the fair outputs will be delivered in practice. Only a small percentage of users may arrive over a given time window and those users may be ones to whom fewer sensitive items are assigned. The carefully balanced set of recommendation lists may never reach its intended audience. 
Only by tracking actual fairness outcomes over time can a system adapt to the uncertainties of user arrivals and item availabilities that characterize recommendation delivery in practice.\footnote{Only in the specific case of push-type delivery (for example, an email of recommendations sent out to all users) can batch-type fairness solutions meet their stated targets. This is an important type of recommendation but not the whole space of recommender system use cases.}

Our recommendation allocation problem also has some similarities with those found in computational advertising, where specific messages are matched with users in a personalized way~\citep{wang2017display,yuan2012internet}. Because advertising is a paid service, these problems are typically addressed through mechanisms of monetary exchange, such as auctions. There is no counterpart to budgets or bids in our context, which means that solutions in this space do not readily translate to supporting fair recommendation \citep{optimalbiding,edelman2007internet,yuan2013real}. 

In fairness-aware recommendation, both \citet{zehlike2022fair} and \citet{sonboli2020opportunistic} present examples of reranking with multiple protected groups. Both are static solutions, without the adaptive capabilities that we seek here. Also, the solution in \cite{zehlike2022fair} depends on a computationally-intensive pre-processing step and cannot be easily adapted to a dynamic setting. We draw from \cite{sonboli2020opportunistic} in our definition of compatibility between an agent and a recommendation opportunity. 

\section{SCRUF-D platform}
SCRUF-D \cite{aird2023dynamic} (and its predecessor SCRUF \cite{sonboli2020and}) are recommendation architectures for integrating fairness into recommendation generation. Both variants of SCRUF can be understood as a form of dynamic recommendation reranking, one of the most common approaches for fairness-aware recommendation \cite{ekstrand2022fairness}, since a recommendation list from a base recommendation algorithm is one of its inputs.

{\setlength\textfloatsep{10pt}
\begin{figure}[tb]
        \centering
        \includegraphics[scale=0.40]{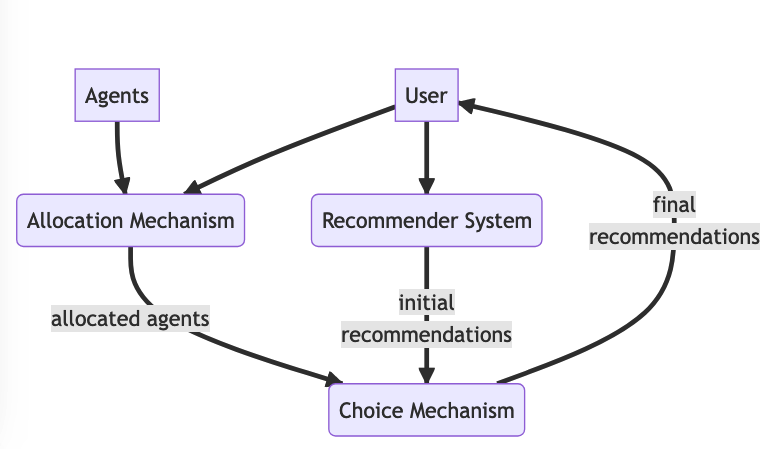}
    \caption{Overview of the SCRUF architecture}
    \label{fig:scruf}
\end{figure}
}

Figure~\ref{fig:scruf} shows an overview of the architecture. The first phase of SCRUF's operation allocates agents to recommendation opportunities (i.e. user arrivals). Only allocated agents can participate in the subsequent choice (voting) phase and have an impact on the generated recommendations. In the second phase, the recommender system and the allocated agent(s) cast ballots, i.e., a ranking / scoring of items, and a preference aggregation mechanism (voting rule) combines them to produce the final list.\footnote{Agents can, in principle, generate their rankings over any set of items but in our experiments so far, we have restricted them to constructing preferences only over those items that the recommender system has returned. } 

To achieve the allocation, the mechanism takes into account two aspects of the current recommendation context: fairness and compatibility. In measuring fairness, each agent tracks the level of fairness achieved over a historical time window, relative to an individually-defined fairness metric. Historical tracking of the state of fairness gives the model its dynamic character, enabling the system to respond to unfairness generated by a particular sequence of user arrivals. In the compatibility function, each agent also measures the expected propensity of the user to respond to recommendations of sensitive items within that agent's purview. This capability corresponds to the notion of \textit{personalized fairness} outlined in \cite{liu2018personalizing,sonboli2020opportunistic}, where the application of a fairness intervention is tailored to each user's historical profile. The distinction between fairness (effectively the agent's need to be allocated) and  compatibility (the likely interest of the user in the items the agent promotes) makes the allocation problem an online capacitated two-sided matching problem \cite{Manlove:MatchingPrefs,lian2018conference}. This formulation assumes that agents' fairness and compatibility metrics are comparable. We are working to develop standard formalizations of these metrics for a wide class of possible fairness metrics.

\section{Mechanisms}
SCRUF-D provides a general framework in which the properties of different allocation and aggregation mechanisms can be explored \cite{aird2023dynamic}. In this paper, we experiment with combinations of mechanisms using simulated and real-world data.

\subsection{Allocation Mechanisms}

We explore three allocation mechanisms using different logics to allocate agents to recommendation opportunities.\\
\textbf{Least Fair:} Under \textit{Least Fair}, the fairness agent with the lowest fairness score is chosen. This mechanism ignores compatibility and focuses on the agent most in need of improvement. However, this approach can lead to starvation -- if the fairness of an agent is slow to improve, it will continue to be allocated and other agents will get no opportunities to achieve their preferred outcomes.\\
\textbf{Lottery:} Our \textit{Lottery} mechanism selects a single fairness agent in the allocation phase using a lottery computed over all agents. We compute the product of unfairness and the square of the compatibility for each agent, normalized to a sum of 1, and then draw an agent from this distribution.\footnote{The tunable exponent for the compatibility term reduces its influence in the lottery. This can be adjusted to favor accuracy over fairness.} This avoids the problem of starvation (since the choice is not deterministic) and factors in compatibility.\\
\textbf{Weighted:} Under \textit{Weighted} allocation, all agents are allocated but their resulting weight is determined by the product between unfairness and (squared) compatibility as in the case of the Lottery method above. This allocation method generates the most complex choice problem since all agents participate in the aggregation phase.

For all allocation mechanisms, the fairness agent definition is used to calculate the fairness scores. In the experiments here, agents have given the same fairness metrics, so the fairness scores used in these allocation mechanisms are directly comparable. However, even when the metrics are not identical across agents, their standardized form means that that they can be compared for allocation purposes. 


\subsection{Choice Mechanisms (Voting Rules)}
We examine four different choice mechanisms. In computational social choice, choice mechanisms are classically understood as integrating the preferences of multiple agents together to form a single societal preference \cite{BCELP16a}.\footnote{Our setting differs from classical social choice in that voting is not anonymous (the recommender system plays a different role from the other agents) and weights and scores are typically employed. Typically, rankings are preferred because they do not require agent utility to be known or knowable.} 
\\
\textbf{Rescoring:} The simplest mechanism is one in which each agent contributes a weighted score for each item and these scores are summed to determine the rank of items. Each fairness agent has a fixed score increment $\delta$ that is added to all protected items, weighted by its allocation in the previous phase. This is combined with the scoring of the recommendation algorithm. \\
\textbf{Borda:} Under the Borda mechanism \cite{DBLP:reference/choice/Zwicker16}, ranks are associated with scores and the original scores used to compute those ranks are ignored. The ranks across agents are summed and the result determines the final ranking. \\
\textbf{Copeland:} The Copeland mechanism calculates a win-loss record for each item considering all item-item pairs in a graph induced by the preferences. Item $i$ scores one point over item $j$ if the majority of allocated agents prefer $i$ to $j$. We then sum these pairwise match-ups for each item $i$ and order the list of items using these scores \cite{sep-voting-methods}. \\
\textbf{RankedPairs:} The Ranked Pairs voting rule \cite{tideman1987independence} computes the pairwise majority graph as described for Copeland but orders the resulting ranking by how much a particular item wins by, selecting these in order to create a complete ranking, skipping a pair if and only if it would induce a cycle in the aggregate ranking. 

Each of these choice mechanisms implements a fundamentally different logic for aggregating preferences: score-based, ordinal-based, consistency-based and pairwise-preference \cite{DBLP:reference/choice/Zwicker16}. As we show in our results, choice mechanisms yield quite different accuracy / fairness tradeoffs.

\section{Methodology}
To explore our research questions, we conducted experiments with both synthetic and real-world data and compared different combinations of mechanisms. Because our research is exploring aspects of these mechanisms, it was not necessary to explore different base recommendation algorithms. For the simulated experiments, we generated synthetic recommender system output as described below. For Microlending, our real-world example, we used a simple biased matrix factorization technique. We determined in prior experiments that this algorithm suffers from popularity bias and therefore represents a challenge for fairness-aware reranking.

\subsection{SCRUF-D Implementation}
SCRUF-D is implemented in Python and available open-source from GitHub under the MIT License.\footnote{https://github.com/that-recsys-lab/scruf\_d} SCRUF-D integrates Whalrus \footnote{https://francois-durand.github.io/whalrus/} for the implementation of choice mechanisms. 
The configuration files, data and analysis code used for our experiments are also available along with the source code for our synthetic data generator. \footnote{https://github.com/that-recsys-lab/SCRUF\_FAccTRec\_2023}


\subsection{Synthetic Data}
The purpose of synthetic data in our simulations is to supply realistic recommender system output as input to the SCRUF-D reranker. 
We create synthetic data via latent factor generation: we create matrices of latent factors similar to those that would be created through factorization and then generate sample ratings from these matrices. Let $\hat{U}$ and  $\hat{V}$ be the user and item latent factor matrices with $k$ latent factors. We designate the first $k_s$ of the latent factors as corresponding to protected features of items, and the remaining $k - k_s$ factors correspond to other aspects of the items.

As a first step, we generate a vector of real-valued propensities for each user $\Phi_i = \langle\phi_1, ..., \phi_{k_s}\rangle$ corresponding to the sensitive features plus additional values for each of the non-sensitive features, drawn from an experimenter-specified normal distribution. Thus, it is possible to adjust the preferences of the user population regarding different sensitive features. The propensities associated with a sensitive feature also represent the user's compatibility with the respective fairness agent, a value which in a non-synthetic case is derived from the pre-existing user profile as in \cite{sonboli2020opportunistic}. 

From $\Phi_i$, we perform an additional generation step to draw a latent factor vector $U_i$ from a normal distribution centered on the propensities. This two-step process avoids having the latent factors tied exactly to the user propensities, which would otherwise make the compatibility of users with agents highly deterministic. 

The profiles for items are generated in a similar way except that items have a binary association with their associated sensitive features and so the experimenter input consists of parameters for a multi-variate Bernoulli distribution. Each item's propensity is generated as a binary vector $\Phi_j$ using these probabilities. As with users, there is a second step of latent factor generation, in which the elements of an item's latent feature vector $V_j$ is drawn from a normal distribution centered on the item's (binary) propensity for that feature. This two-step procedure allows us to identify an item as possessing a particular feature (particularly protected ones) without the latent factor encoding this exactly. 

After $\hat{U}$ and $\hat{V}$ have been generated, we then select $m$ items at random for each user $i$ and compute the product of the $\hat{U}_i$ and $\hat{V}_j$ as the synthetic rating for each user $i$, item $j$ pair. We sort these values and select the top $m'$ as the recommender system output. The sorting / filtering process ensures that the output is biased towards more highly-rated items, which is what one would expect in recommender system output.

For the first set of experiments in this paper, we generated 500 users and 5,000 items. For each user, we generated 200 sample ratings and used the top 50 as the recommendation lists. Item propensities were set to 0.039, 0.05, 0.9 for three latent factors; the third factor being not sensitive. The standard deviation of the factors was 1.0. Corresponding user propensities for the features were $\mu = 0.5, \sigma = 0.06$ for the first protected feature and $\mu = 0.6, \sigma = 0.08$ for the second feature. The generation parameters were based on proportions seen in real-world datasets including the Microlending dataset described below. This dataset is referred to as \textit{Synthetic} in the discussion below.

To explore dynamic aspects of the system's responses, we created additional datasets similar to the one described above but where the users are split into different three segments $<A, B, C>$, each arriving in sequence. We generated synthetic users with high compatibility to Agent 2 and low compatibility with Agent 1 in segment $A$, then reversed this affinity in segment $B$. Segment $C$ contained used without high compatibility with either agent. 
We used different generating parameters than above, making the differences between user types more extreme and with a lower prevalence of protected items. This data is referred to as $Synthetic Sequenced$.

\subsection{Microlending Data}
In addition to the Synthetic data, we used the Microlending 2017 dataset \cite{sonboli2022micro}, which contains anonymized lending transactions from the crowd-sourced microlending site Kiva.org. The dataset has 2,673 pseudo-items, 4,005 lenders and 110,371 ratings / lending actions. See \cite{sonboli2020opportunistic} and \cite{sonboli2022micro} for a complete description of the data set. 

We considered two loan feature categories, loan size and country, as protected features. Prior work \cite{sonboli2020opportunistic} identified loan size as a dimension along which fairness in lending may need to be sought. About 4\% of loans had this feature and were considered protected items. For the second protected feature, we followed \citet{sonboli2020opportunistic} in identifying the 16 countries whose loans have the lowest rates of funding  and labeled these as the protected group for the purposes of geographic fairness. Compatibility scores were defined using the entropy of a user's ratings versus the protected status of funded loans using the method in \cite{sonboli2020opportunistic}. 

We were not able to duplicate the conditions in the synthetic data because there is a high correlation between users' entropies with respect to the country variable and with respect to the loan size variable. 
In other words, users who were highly compatible with the loan size agent were also for the most part also compatible with the country agent. So, it was not possible to have segments of users with differential compatibility for different agents and we used only synthetic data for looking at different orderings of users.

\subsection{Agent definitions}
For these experiments, we assume a single fairness definition, that of group proportional fairness: a fixed desirable proportion $\pi$ of protected group items in each recommendation list. To calculate overall fairness, we create a union of all recommender lists within the history window and calculate the proportion of agent-specific protected items. We scale this proportion dividing by $\pi$ and truncate larger values at 1.0. 

For the Synthetic data, we set the fairness target for both Agents 1 and 2 to be 25\%. 
In Microlending, we set the target for the loan size agent to be 30\% and for the country agent to 20\%. Fairness agents were allocated using scores based on these target proportions. 

When a fairness agent is allocated, the scores from the recommender systems are adjusted such that each item among the protected items has its score augmented by a constant $\delta$. For the Synthetic data, $\delta$ was set at 0.1 for both agents. For Microlending, $\delta$ was set at 0.3 for the country agent and 0.6 for the loan size agent to give more impact to the loan size agent, as these items were more difficult to reach the fairness proportion. 



\section{Results}
For the Microlending experiments, we generated 50 recommendations for each of the 4,005 users using biased matrix factorization as implemented in the LibRec recommendation library \footnote{https://github.com/guoguibing/librec}. We chose this recommendation algorithm as input for these experiments as it has known issues with popularity bias \cite{jannach2015recommenders} and therefore presents a challenging reranking problem: as noted above, any recommendation algorithm could be used. We generated similar recommendation lists for the synthetic data as noted above. In both experiments, we set the history window, over which agents consider fairness outcomes, to be 100 users. 

In reporting results, we calculate fairness by normalizing the proportion of protected items for each agent across the whole experiment and averaging across the agents. (Note that this is different than the window-limited metric that agents have access to.) We use a proportional fairness measure in these experiments, following \cite{sonboli2020opportunistic}, where recommendation list exposure measured in this way was the key metric and this is also the measure used by the agents themselves. We use nDCG@10 throughout for recommendation accuracy. However, for the synthetic data sets, there is no ground truth test data against which to compute utility, so for these experiments we compute nDCG relative to the original recommendation lists.


Figure \ref{fig:scatter} summarizes the results from the experiments with the randomly sequenced users, plotting accuracy vs fairness. There are clear groupings of the choice mechanisms at different tradeoff points. In the Synthetic data, the order (from most accurate / least fair to least accurate / most fair) is (roughly) Ranked Pairs, Rescoring, Borda and Copeland, although Ranked Pairs is dominated by Rescoring / Weighted. For the Microlending data, the order is Rescoring, Borda, Ranked Pairs and Copeland with Borda dominated by Rescoring / Lottery.

{\setlength\textfloatsep{10pt}
\begin{figure}[bt]
   \centering
    \begin{subfigure}[b]{0.48\textwidth}
        \centering
        \includegraphics[scale=0.40]{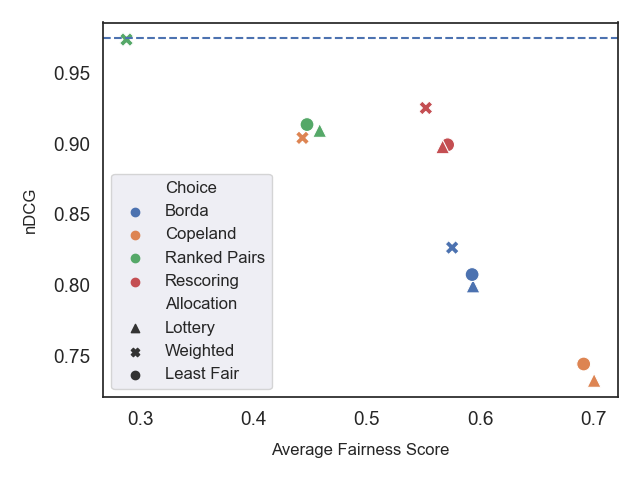}
        \caption{Synthetic Data}
        \label{fig:scatter-synthetic}
    \end{subfigure}
    \hfill
    \begin{subfigure}[b]{0.48\textwidth}
        \centering
        \includegraphics[scale=0.40]{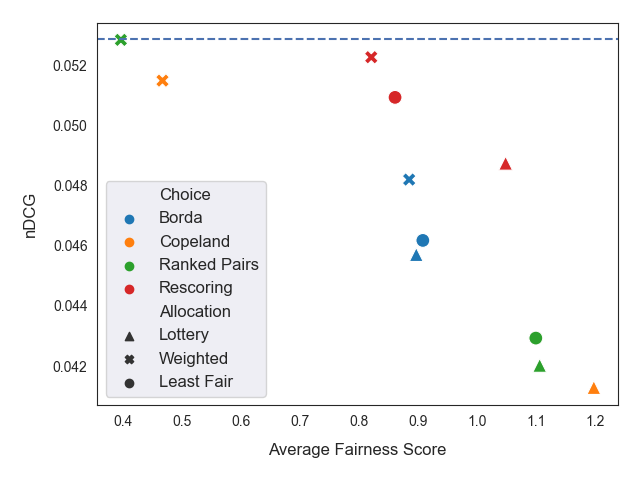}
        \caption{Microlending Data}
        \label{fig:scatter-microlending}
    \end{subfigure}
    \caption{Accuracy vs average normalized fairness. Fairness target is at 1.0; baseline accuracy is shown in the dashed line.}
    \label{fig:scatter}
\end{figure}
}

Weighted allocation breaks this pattern in interaction with the two concordance-based mechanisms. The Weighted agents have individually lower weights and there is rarely any synergy between them, so they are effectively outvoted by the recommender. Although Ranked Pairs and Copeland are both concordance-based methods, they have different methods of handling ties (partial orders) and because our fairness agents produce partial orders, the tie-breaking method is significant. Copeland effectively scores a tie as 0.5 of a concordant pair, while Ranked Pairs breaks ties randomly and a much larger set of possible orderings can arise. Many of these orderings are ones in which the recommender's rankings dominate. The effect is not nearly as pronounced with the Borda and Rescoring mechanisms because the output from the multiple agents are combined in an additive way.

The two other allocation mechanisms, Lottery and Least Fair, are surprisingly similar in their outcomes. (And, in fact, identical and superimposed for the Copeland mechanism in the Microlending dataset.) We expected that Least Fair would have lower accuracy since it ignores user compatibility in assigning agents. But with a few exceptions, across the different conditions and datasets, we find that the Lottery mechanism is associated with lower accuracy and greater fairness. Interestingly, the loss of accuracy is largest for the Rescoring mechanism and the Kiva dataset and smallest for Copeland in the Kiva dataset where the data points are superimposed. What we do see if we examine the individual agent scores is that Least Fair is associated with a greater difference in fairness results across agents so the difference may have to do a certain amount of starvation occurring in the experiments.

Across the two datasets, one striking difference is in the performance of Ranked Pairs. It is relatively ineffective (and Pareto-dominated) in the Synthetic data and very close to Copeland in the Kiva data. One possible reason again traces back to the differences in ranking processes. The Synthetic data is fairly simple in structure and lower in noise. In a higher entropy dataset, the differences in tie-breaking procedures do not have as much of an impact. 

Figure~\ref{fig:box-compare} shows the distribution of each agent's fairness metric as computed in each time interval. The baseline (non-reranked) fairness values are shown as the dashed lines, and we see that the baseline fairness is greater for Agent 1. In most cases, the two agents are fairly close in average fairness across the time steps of the experiment, showing that the mechanisms are working together to equalize the agents' outcomes. Some conditions actually favor Agent 2, particularly Least Fair + Rescoring, Lottery + Copeland, and the low fairness result of Weighted + Copeland.\footnote{Because of the correlation between agent compatibilities in the Microlending data, there is very little difference between the fairness results across agents and so we do not include a similar plot for those experiments.}

{\setlength\textfloatsep{10pt}
\begin{figure}[bt]
   \centering
    \includegraphics[scale=0.38]{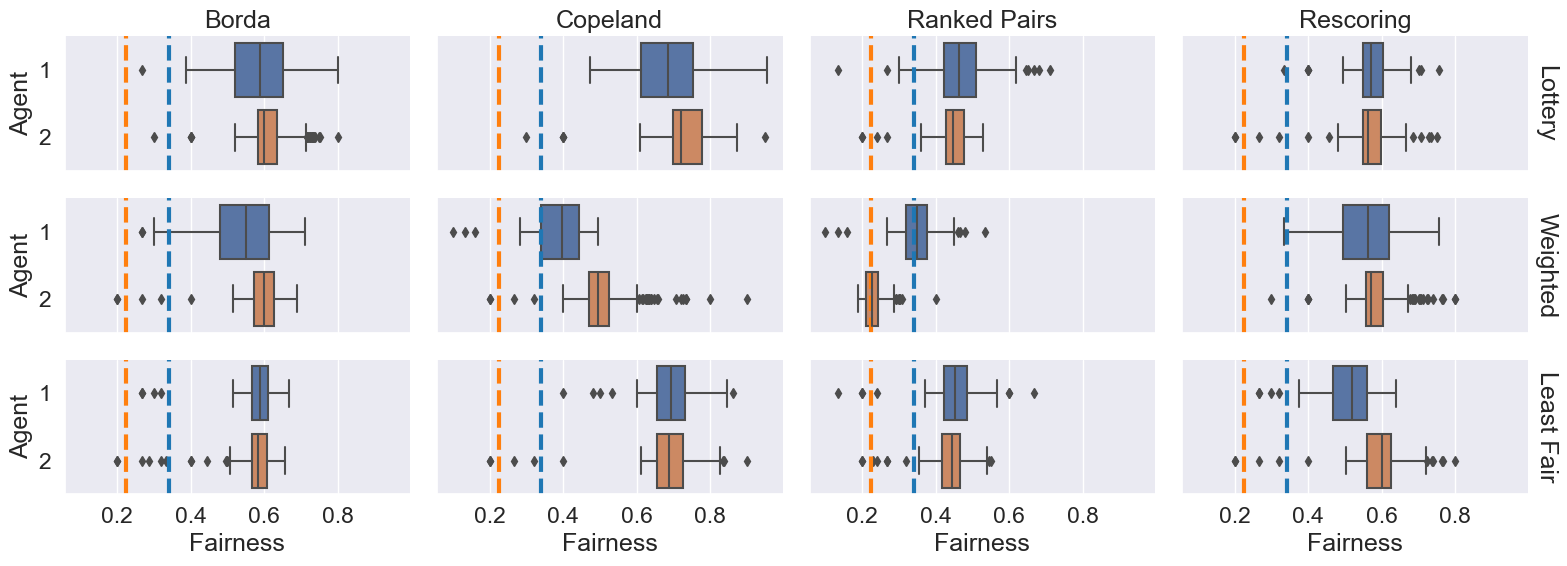}
    \caption{Fairness metric distribution for each agent (Synthetic Data)}
    \label{fig:box-compare}
\end{figure}
}

The second set of experiments are those in which the users were segmented into different regimes as described above. In these experiment, we focus on the ability of the different mechanisms to cope with changes in relative abundance of compatible users. Because the data was generated in a different way, the Synthetic Sequenced data is not comparable to the original Synthetic data. 

{\setlength\textfloatsep{10pt}
\begin{figure}[tb]
        \centering
        \includegraphics[scale=0.66]{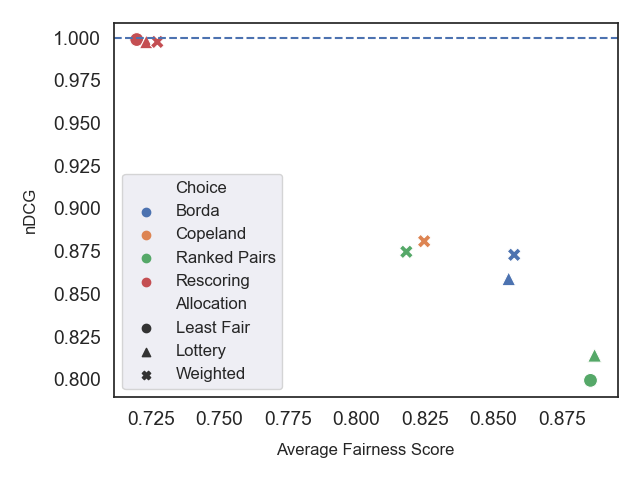}
    \caption{Accuracy vs average normalized fairness in segmented experiment. Fairness target is at 1.0; baseline accuracy is shown in the dashed line.}
    \label{fig:scatter-segmented}
\end{figure}
}


Figure~\ref{fig:scatter-segmented} shows the results from these dynamic experiments. Recall that both agents' fairness targets are more difficult in the Synthetic Sequenced data set. The result here are more similar to what we saw with the Microlending data above. Ranked Pairs occupies the low accuracy / high fairness area and Rescoring, the high accuracy upper left. Note that that fairness here is still an improvement over the baseline (at 0.65). The Least Fair methods all occupy more or less the same overlapping position in this corner, likely because they tend to have worse fairness results for Agent 0, as evidenced in the allocation plots below. We see that the Weighted allocation is more often a good option with this dataset, most likely because it allocates all the agents in every iteration and includes compatibility in its weighting.
 
{\setlength\textfloatsep{10pt}
\begin{figure}[tb]
   \centering
    \begin{subfigure}[b]{0.45\textwidth}
        \centering
        \includegraphics[scale=0.25]{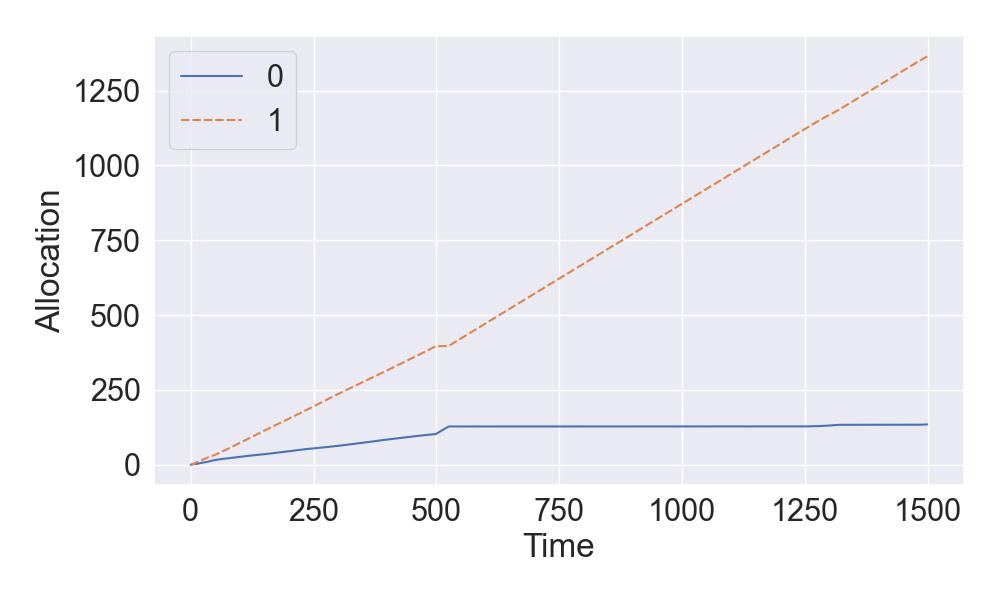}
        \caption{Least Fair Allocation}
        \label{fig:scatter-synthetic-segmented}
    \end{subfigure}
    \hfill
    \begin{subfigure}[b]{0.45\textwidth}
        \centering
        \includegraphics[scale=0.25]{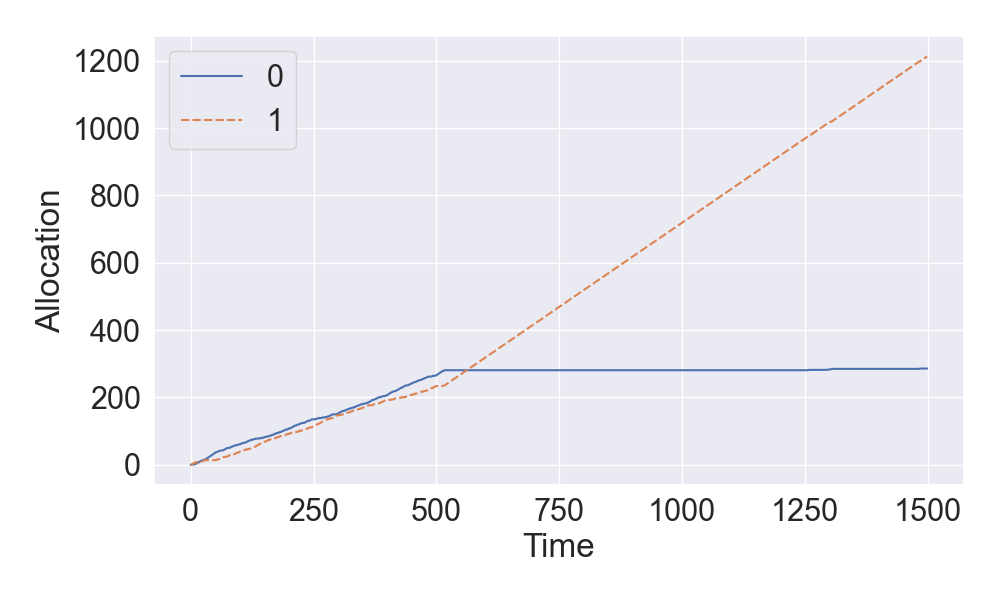}
        \caption{Weighted Allocation}
        \label{fig:dynamic_allocation}
    \end{subfigure}
    \caption{Cumulative allocation of fairness agents with different allocation mechanisms}
    \label{fig:dynamic-allocation}
\end{figure}
}

In Figure~\ref{fig:dynamic_allocation}, we see the contrast between allocation mechanisms in the Synthetic Sequenced where there are strong differences between types of users over time. Both examples use Ranked Pairs as the choice mechanism. The Least Fair mechanism keeps trying to allocate Agent 1 because it is more difficult to achieve fairness for this agent, even though the initial set of users is not very compatible with this agent's fairness objective. Agent 0 is relatively starved as a result. The Weighted allocation includes both agents in its allocation for the first set of users taking advantage of the opportunity presented by the compatible users in the initial segment. In the end, greater fairness is achieved with the Least Fair mechanism but at substantial cost of accuracy because the preferences of the first segment of users is generally ignored.

\section{Conclusion}
In this paper, we explored combinations of allocation and choice mechanisms for integrating multiple fairness concerns into recommendation. Relative to RQ1, we find that there are consistent differences between combinations of mechanisms, placing them at different points along a fairness-accuracy frontier. Although the ranking is not completely consistent across datasets, Copeland is generally towards the bottom right, except when Weighted allocation is used and the agents have less impact. Rescoring occupies a lower fairness, higher accuracy position. Borda is in between. The Synthetic Sequenced data, in which protected items were more rare, looks more like the Microlending results, suggesting that the ``difficulty'' of the fairness problem impacts the relative efficacy of the different mechanisms. We will explore this phenomenon further in our future work.

Considering RQ3, allocation mechanisms have a smaller impact on this tradeoff in most cases, the exception being the Weighted allocation, which interacts with the pair-wise methods in a manner that greatly reduces the impact in both fairness and accuracy dimensions. Across experiments, Weighted, Least Fair and Lottery are generally ranked in that order along the high accuracy / high fairness tradeoff diagonal. 

Our segmented experiments addressed RQ 2. We see that in most cases, the Least Fair mechanism is unable to improve fairness beyond a certain point because it is blind to user-agent compatibility and as a result, fails to take advantage of recommendation opportunities as completely as some of the other methods. We also saw that there were benefits to the Weighted mechanism that were not as apparent in the randomly ordered data. These findings will get further investigation in our future work.

This paper represents the first evaluation of the fairness / accuracy tradeoff of multiple fairness objectives using a social choice framework and as such it is very preliminary. The phenomena found here need to be explored in much greater detail. We plan to vary the properties of the synthetic data and incorporate additional real datasets, explore a range of fairness targets, and investigate the dynamics of the system more fully. We also intend to compare our results with alternative algorithms for multigroup fairness-aware reranking, especially OFAiR \cite{sonboli2020opportunistic}.

There are three key areas that we think it is crucial to address in future research. The first is the issue of multiple fairness definitions. In this study, all of the fairness agents use the same fairness definition and metric. That is typical of fairness-aware recommendation research, although not typical of practical fairness settings \cite{smith2023many}. It will be important to explore how SCRUF operates in the presence of different fairness metrics and logics, including non-binary and continuous definitions and consumer-side fairness \cite{ekstrand2022fairness}. A consumer-side fairness definition could use a fairness agent that is concerned in the cost of fairness, in terms of accuracy, to users. We can think of this as the difference between what the recommender would have returned without the fairness intervention and what actually was returned, although of course there is no great certainty that the recommender system always represents users' preferences well. Rank correlation or a similar measure can be used to compare these lists. The evaluation metric would focus on variance of this statistic. If all users are experiencing relatively similar accuracy losses, the metric will have a low value. 

As we expand the scope of fairness considerations, we will inevitably find ourselves in a situation with a larger collection of agents than the 2 deployed in these experiments. While we do not expect that applications will need unbounded numbers of agents, our research suggests that between 5-10 will be needed in the Kiva context. Additional research is needed to examine the characteristics of larger agent collections. However, one of the key advantages of social choice mechanisms is that they are designed to handle multiple agents in interaction, so we expect that key findings for smaller numbers of agents will extend to larger groups.

A second research direction is in the representation of preferences. Social choice methods (including all of those used here except Rescoring) are based on ranks and do not take scores into consideration. To incorporate a quantitative aspect, while sticking with the social choice formalization, we can represent recommendation outputs as \textit{stochastic rankings} \cite{diaz2020evaluating} from which it is possible to generate multiple different ballots. 

Lastly, we note the assumption in this work that the balance between the fairness agents and the recommender system can be represented statically as a single parameter of recommender system weight. If this weight is too high, the agents do not have the opportunity to influence rankings at all (particularly under the winner-take-all concordance mechanisms). If the weight is too low, accuracy will potentially be unacceptably low. In future work, we will explore the use of bandit mechanisms to establish appropriate policies for striking this balance. We will also explore how this may change as more agents are added. 


\section*{Acknowledgements}
This research was supported by the National Science Foundation under awards IIS-2107577 and IIS-2107505. This work was also supported by Slovak Research and Development Agency under Contract no. APVV-20-0353 and the Fulbright Program.

\bibliographystyle{ACM-Reference-Format}
\bibliography{bib}


\begin{thebibliography}{30}


\ifx \showCODEN    \undefined \def \showCODEN     #1{\unskip}     \fi
\ifx \showDOI      \undefined \def \showDOI       #1{#1}\fi
\ifx \showISBNx    \undefined \def \showISBNx     #1{\unskip}     \fi
\ifx \showISBNxiii \undefined \def \showISBNxiii  #1{\unskip}     \fi
\ifx \showISSN     \undefined \def \showISSN      #1{\unskip}     \fi
\ifx \showLCCN     \undefined \def \showLCCN      #1{\unskip}     \fi
\ifx \shownote     \undefined \def \shownote      #1{#1}          \fi
\ifx \showarticletitle \undefined \def \showarticletitle #1{#1}   \fi
\ifx \showURL      \undefined \def \showURL       {\relax}        \fi
\providecommand\bibfield[2]{#2}
\providecommand\bibinfo[2]{#2}
\providecommand\natexlab[1]{#1}
\providecommand\showeprint[2][]{arXiv:#2}

\bibitem[Aird et~al\mbox{.}(2023)]%
        {aird2023dynamic}
\bibfield{author}{\bibinfo{person}{Amanda Aird}, \bibinfo{person}{Paresha
  Farastu}, \bibinfo{person}{Joshua Sun}, \bibinfo{person}{Amy Voida},
  \bibinfo{person}{Nicholas Mattei}, {and} \bibinfo{person}{Robin Burke}.}
  \bibinfo{year}{2023}\natexlab{}.
\newblock \bibinfo{title}{Dynamic fairness-aware recommendation through
  multi-agent social choice}.
\newblock
\newblock
\showeprint[arxiv]{2303.00968}~[cs.AI]


\bibitem[Brandt et~al\mbox{.}(2016)]%
        {BCELP16a}
\bibfield{editor}{\bibinfo{person}{F. Brandt}, \bibinfo{person}{V. Conitzer},
  \bibinfo{person}{U. Endriss}, \bibinfo{person}{J. Lang}, {and}
  \bibinfo{person}{A.~D. Procaccia}} (Eds.). \bibinfo{year}{2016}\natexlab{}.
\newblock \bibinfo{booktitle}{\emph{Handbook of Computational Social Choice}}.
\newblock \bibinfo{publisher}{Cambridge University Press}.
\newblock


\bibitem[Burke(2017)]%
        {burke_multisided_2017}
\bibfield{author}{\bibinfo{person}{Robin Burke}.}
  \bibinfo{year}{2017}\natexlab{}.
\newblock \showarticletitle{Multisided {Fairness} for {Recommendation}}. In
  \bibinfo{booktitle}{\emph{Workshop on {Fairness}, {Accountability} and
  {Transparency} in {Machine} {Learning} ({FATML})}}.
  \bibinfo{address}{Halifax, Nova Scotia}, \bibinfo{numpages}{5}~pages.
\newblock
\urldef\tempurl%
\url{https://arxiv.org/abs/1707.00093}
\showURL{%
\tempurl}


\bibitem[Burke et~al\mbox{.}(2022)]%
        {burke2022multi}
\bibfield{author}{\bibinfo{person}{Robin Burke}, \bibinfo{person}{Nicholas
  Mattei}, \bibinfo{person}{Vladislav Grozin}, \bibinfo{person}{Amy Voida},
  {and} \bibinfo{person}{Nasim Sonboli}.} \bibinfo{year}{2022}\natexlab{}.
\newblock \showarticletitle{Multi-agent Social Choice for Dynamic
  Fairness-aware Recommendation}. In \bibinfo{booktitle}{\emph{Adjunct
  Proceedings of the 30th ACM Conference on User Modeling, Adaptation and
  Personalization}}. \bibinfo{pages}{234--244}.
\newblock


\bibitem[Diaz et~al\mbox{.}(2020)]%
        {diaz2020evaluating}
\bibfield{author}{\bibinfo{person}{Fernando Diaz}, \bibinfo{person}{Bhaskar
  Mitra}, \bibinfo{person}{Michael~D Ekstrand}, \bibinfo{person}{Asia~J Biega},
  {and} \bibinfo{person}{Ben Carterette}.} \bibinfo{year}{2020}\natexlab{}.
\newblock \showarticletitle{Evaluating stochastic rankings with expected
  exposure}. In \bibinfo{booktitle}{\emph{Proceedings of the 29th ACM
  international conference on information \& knowledge management}}.
  \bibinfo{pages}{275--284}.
\newblock


\bibitem[Edelman et~al\mbox{.}(2007)]%
        {edelman2007internet}
\bibfield{author}{\bibinfo{person}{Benjamin Edelman}, \bibinfo{person}{Michael
  Ostrovsky}, {and} \bibinfo{person}{Michael Schwarz}.}
  \bibinfo{year}{2007}\natexlab{}.
\newblock \showarticletitle{Internet advertising and the generalized
  second-price auction: Selling billions of dollars worth of keywords}.
\newblock \bibinfo{journal}{\emph{The American economic review}}
  \bibinfo{volume}{97}, \bibinfo{number}{1} (\bibinfo{year}{2007}),
  \bibinfo{pages}{242--259}.
\newblock


\bibitem[Ekstrand et~al\mbox{.}(2022)]%
        {ekstrand2022fairness}
\bibfield{author}{\bibinfo{person}{Michael~D Ekstrand},
  \bibinfo{person}{Anubrata Das}, \bibinfo{person}{Robin Burke},
  \bibinfo{person}{Fernando Diaz}, {et~al\mbox{.}}}
  \bibinfo{year}{2022}\natexlab{}.
\newblock \showarticletitle{Fairness in information access systems}.
\newblock \bibinfo{journal}{\emph{Foundations and Trends{\textregistered} in
  Information Retrieval}} \bibinfo{volume}{16}, \bibinfo{number}{1-2}
  (\bibinfo{year}{2022}), \bibinfo{pages}{1--177}.
\newblock


\bibitem[Freeman et~al\mbox{.}(2017)]%
        {freeman2017fair}
\bibfield{author}{\bibinfo{person}{Rupert Freeman},
  \bibinfo{person}{Seyed~Majid Zahedi}, {and} \bibinfo{person}{Vincent
  Conitzer}.} \bibinfo{year}{2017}\natexlab{}.
\newblock \showarticletitle{Fair social choice in dynamic settings}. In
  \bibinfo{booktitle}{\emph{Proceedings of the 26th International Joint
  Conference on Artificial Intelligence (IJCAI)}}.
  \bibinfo{publisher}{International Joint Conferences on Artificial
  Intelligence}, \bibinfo{address}{Marina del Rey, CA},
  \bibinfo{pages}{4580--4587}.
\newblock


\bibitem[Ge et~al\mbox{.}(2021)]%
        {ge2021towards}
\bibfield{author}{\bibinfo{person}{Yingqiang Ge}, \bibinfo{person}{Shuchang
  Liu}, \bibinfo{person}{Ruoyuan Gao}, \bibinfo{person}{Yikun Xian},
  \bibinfo{person}{Yunqi Li}, \bibinfo{person}{Xiangyu Zhao},
  \bibinfo{person}{Changhua Pei}, \bibinfo{person}{Fei Sun},
  \bibinfo{person}{Junfeng Ge}, \bibinfo{person}{Wenwu Ou}, {et~al\mbox{.}}}
  \bibinfo{year}{2021}\natexlab{}.
\newblock \showarticletitle{Towards long-term fairness in recommendation}. In
  \bibinfo{booktitle}{\emph{Proceedings of the 14th ACM International
  Conference on Web Search and Data Mining}}. \bibinfo{publisher}{ACM},
  \bibinfo{address}{New York}, \bibinfo{pages}{445--453}.
\newblock


\bibitem[Jannach et~al\mbox{.}(2015)]%
        {jannach2015recommenders}
\bibfield{author}{\bibinfo{person}{Dietmar Jannach}, \bibinfo{person}{Lukas
  Lerche}, \bibinfo{person}{Iman Kamehkhosh}, {and} \bibinfo{person}{Michael
  Jugovac}.} \bibinfo{year}{2015}\natexlab{}.
\newblock \showarticletitle{What recommenders recommend: an analysis of
  recommendation biases and possible countermeasures}.
\newblock \bibinfo{journal}{\emph{User Modeling and User-Adapted Interaction}}
  \bibinfo{volume}{25} (\bibinfo{year}{2015}), \bibinfo{pages}{427--491}.
\newblock


\bibitem[Lackner(2020)]%
        {lackner2020perpetual}
\bibfield{author}{\bibinfo{person}{Martin Lackner}.}
  \bibinfo{year}{2020}\natexlab{}.
\newblock \showarticletitle{Perpetual voting: Fairness in long-term decision
  making}. In \bibinfo{booktitle}{\emph{Proceedings of the AAAI conference on
  artificial intelligence}}, Vol.~\bibinfo{volume}{34}.
  \bibinfo{pages}{2103--2110}.
\newblock


\bibitem[Lian et~al\mbox{.}(2018)]%
        {lian2018conference}
\bibfield{author}{\bibinfo{person}{Jing~Wu Lian}, \bibinfo{person}{Nicholas
  Mattei}, \bibinfo{person}{Renee Noble}, {and} \bibinfo{person}{Toby Walsh}.}
  \bibinfo{year}{2018}\natexlab{}.
\newblock \showarticletitle{The conference paper assignment problem: Using
  order weighted averages to assign indivisible goods}. In
  \bibinfo{booktitle}{\emph{Proceedings of the AAAI Conference on Artificial
  Intelligence}}.
\newblock


\bibitem[Liu and Burke(2018)]%
        {liu2018personalizing}
\bibfield{author}{\bibinfo{person}{Weiwen Liu} {and} \bibinfo{person}{Robin
  Burke}.} \bibinfo{year}{2018}\natexlab{}.
\newblock \showarticletitle{Personalizing Fairness-aware Re-ranking}.
\newblock \bibinfo{journal}{\emph{arXiv preprint arXiv:1809.02921}}
  (\bibinfo{year}{2018}).
\newblock
\newblock
\shownote{Presented at the 2nd FATRec Workshop held at RecSys 2018, Vancouver,
  CA.}.


\bibitem[Manlove(2013)]%
        {Manlove:MatchingPrefs}
\bibfield{author}{\bibinfo{person}{David~F. Manlove}.}
  \bibinfo{year}{2013}\natexlab{}.
\newblock \bibinfo{booktitle}{\emph{Algorithmics of Matching Under
  Preferences}}. \bibinfo{series}{Series on Theoretical Computer Science},
  Vol.~\bibinfo{volume}{2}.
\newblock \bibinfo{publisher}{WorldScientific}.
\newblock
\showISBNx{978-981-4425-24-7}
\urldef\tempurl%
\url{https://doi.org/10.1142/8591}
\showDOI{\tempurl}


\bibitem[Morik et~al\mbox{.}(2020)]%
        {morik2020controlling}
\bibfield{author}{\bibinfo{person}{Marco Morik}, \bibinfo{person}{Ashudeep
  Singh}, \bibinfo{person}{Jessica Hong}, {and} \bibinfo{person}{Thorsten
  Joachims}.} \bibinfo{year}{2020}\natexlab{}.
\newblock \showarticletitle{Controlling fairness and bias in dynamic
  learning-to-rank}. In \bibinfo{booktitle}{\emph{Proceedings of the 43rd
  International ACM SIGIR Conference on Research and Development in Information
  Retrieval}}. \bibinfo{publisher}{ACM}, \bibinfo{address}{New York},
  \bibinfo{pages}{429--438}.
\newblock


\bibitem[Pacuit(2019)]%
        {sep-voting-methods}
\bibfield{author}{\bibinfo{person}{Eric Pacuit}.}
  \bibinfo{year}{2019}\natexlab{}.
\newblock \showarticletitle{{Voting Methods}}.
\newblock In \bibinfo{booktitle}{\emph{The {Stanford} Encyclopedia of
  Philosophy} (\bibinfo{edition}{{F}all 2019} ed.)},
  \bibfield{editor}{\bibinfo{person}{Edward~N. Zalta}} (Ed.).
  \bibinfo{publisher}{Metaphysics Research Lab, Stanford University}.
\newblock


\bibitem[Parkes and Procaccia(2013)]%
        {parkes2013dynamic}
\bibfield{author}{\bibinfo{person}{David Parkes} {and} \bibinfo{person}{Ariel
  Procaccia}.} \bibinfo{year}{2013}\natexlab{}.
\newblock \showarticletitle{Dynamic social choice with evolving preferences}.
  In \bibinfo{booktitle}{\emph{Proceedings of the AAAI conference on artificial
  intelligence}}, Vol.~\bibinfo{volume}{27}. \bibinfo{pages}{767--773}.
\newblock


\bibitem[Patro et~al\mbox{.}(2020)]%
        {patro2020fairrec}
\bibfield{author}{\bibinfo{person}{Gourab~K Patro}, \bibinfo{person}{Arpita
  Biswas}, \bibinfo{person}{Niloy Ganguly}, \bibinfo{person}{Krishna~P
  Gummadi}, {and} \bibinfo{person}{Abhijnan Chakraborty}.}
  \bibinfo{year}{2020}\natexlab{}.
\newblock \showarticletitle{FairRec: Two-Sided Fairness for Personalized
  Recommendations in Two-Sided Platforms}. In
  \bibinfo{booktitle}{\emph{Proceedings of The Web Conference 2020}}.
  \bibinfo{pages}{1194--1204}.
\newblock


\bibitem[Smith et~al\mbox{.}(2023)]%
        {smith2023many}
\bibfield{author}{\bibinfo{person}{Jessie~J Smith}, \bibinfo{person}{Anas
  Buhayh}, \bibinfo{person}{Anushka Kathait}, \bibinfo{person}{Pradeep
  Ragothaman}, \bibinfo{person}{Nicholas Mattei}, \bibinfo{person}{Robin
  Burke}, {and} \bibinfo{person}{Amy Voida}.} \bibinfo{year}{2023}\natexlab{}.
\newblock \showarticletitle{The Many Faces of Fairness: Exploring the
  Institutional Logics of Multistakeholder Microlending Recommendation}. In
  \bibinfo{booktitle}{\emph{Proceedings of the 2023 ACM Conference on Fairness,
  Accountability, and Transparency}}. \bibinfo{pages}{1652--1663}.
\newblock


\bibitem[Sonboli et~al\mbox{.}(2022)]%
        {sonboli2022micro}
\bibfield{author}{\bibinfo{person}{Nasim Sonboli}, \bibinfo{person}{Amanda
  Aird}, {and} \bibinfo{person}{Robin Burke}.} \bibinfo{year}{2022}\natexlab{}.
\newblock \bibinfo{booktitle}{\emph{Microlending 2017 Data Set}}.
\newblock
\urldef\tempurl%
\url{https://doi.org/10.25810/PGJK-RR19}
\showDOI{\tempurl}


\bibitem[Sonboli et~al\mbox{.}(2020a)]%
        {sonboli2020and}
\bibfield{author}{\bibinfo{person}{Nasim Sonboli}, \bibinfo{person}{Robin
  Burke}, \bibinfo{person}{Nicholas Mattei}, \bibinfo{person}{Farzad
  Eskandanian}, {and} \bibinfo{person}{Tian Gao}.}
  \bibinfo{year}{2020}\natexlab{a}.
\newblock \bibinfo{title}{"And the Winner Is...": Dynamic Lotteries for
  Multi-group Fairness-Aware Recommendation}.
\newblock
\newblock
\showeprint[arxiv]{2009.02590}~[cs.IR]


\bibitem[Sonboli et~al\mbox{.}(2020b)]%
        {sonboli2020opportunistic}
\bibfield{author}{\bibinfo{person}{Nasim Sonboli}, \bibinfo{person}{Farzad
  Eskandanian}, \bibinfo{person}{Robin Burke}, \bibinfo{person}{Weiwen Liu},
  {and} \bibinfo{person}{Bamshad Mobasher}.} \bibinfo{year}{2020}\natexlab{b}.
\newblock \showarticletitle{Opportunistic Multi-Aspect Fairness through
  Personalized Re-Ranking}. In \bibinfo{booktitle}{\emph{Proceedings of the
  28th ACM Conference on User Modeling, Adaptation and Personalization}}
  (Genoa, Italy) \emph{(\bibinfo{series}{UMAP '20})}.
  \bibinfo{publisher}{Association for Computing Machinery},
  \bibinfo{address}{New York, NY, USA}, \bibinfo{pages}{239–247}.
\newblock
\showISBNx{9781450368612}
\urldef\tempurl%
\url{https://doi.org/10.1145/3340631.3394846}
\showDOI{\tempurl}


\bibitem[Thomson(2011)]%
        {Thomson:FairRules}
\bibfield{author}{\bibinfo{person}{William Thomson}.}
  \bibinfo{year}{2011}\natexlab{}.
\newblock \showarticletitle{Fair allocation rules}.
\newblock In \bibinfo{booktitle}{\emph{Handbook of Social Choice and Welfare}}.
  Vol.~\bibinfo{volume}{2}. \bibinfo{publisher}{Elsevier},
  \bibinfo{pages}{393--506}.
\newblock


\bibitem[Tideman(1987)]%
        {tideman1987independence}
\bibfield{author}{\bibinfo{person}{T~Nicolaus Tideman}.}
  \bibinfo{year}{1987}\natexlab{}.
\newblock \showarticletitle{Independence of clones as a criterion for voting
  rules}.
\newblock \bibinfo{journal}{\emph{Social Choice and Welfare}}
  \bibinfo{volume}{4} (\bibinfo{year}{1987}), \bibinfo{pages}{185--206}.
\newblock


\bibitem[Wang et~al\mbox{.}(2017)]%
        {wang2017display}
\bibfield{author}{\bibinfo{person}{Jun Wang}, \bibinfo{person}{Weinan Zhang},
  {and} \bibinfo{person}{Shuai Yuan}.} \bibinfo{year}{2017}\natexlab{}.
\newblock \bibinfo{title}{Display Advertising with Real-Time Bidding (RTB) and
  Behavioural Targeting}.
\newblock
\newblock
\showeprint[arxiv]{1610.03013}~[cs.GT]


\bibitem[Yuan et~al\mbox{.}(2012)]%
        {yuan2012internet}
\bibfield{author}{\bibinfo{person}{Shuai Yuan}, \bibinfo{person}{Ahmad~Zainal
  Abidin}, \bibinfo{person}{Marc Sloan}, {and} \bibinfo{person}{Jun Wang}.}
  \bibinfo{year}{2012}\natexlab{}.
\newblock \bibinfo{title}{Internet Advertising: An Interplay among Advertisers,
  Online Publishers, Ad Exchanges and Web Users}.
\newblock
\newblock
\showeprint[arxiv]{1206.1754}~[cs.IR]


\bibitem[Yuan et~al\mbox{.}(2013)]%
        {yuan2013real}
\bibfield{author}{\bibinfo{person}{Shuai Yuan}, \bibinfo{person}{Jun Wang},
  {and} \bibinfo{person}{Xiaoxue Zhao}.} \bibinfo{year}{2013}\natexlab{}.
\newblock \showarticletitle{Real-time bidding for online advertising:
  measurement and analysis}. In \bibinfo{booktitle}{\emph{Proceedings of the
  Seventh International Workshop on Data Mining for Online Advertising}}. ACM,
  \bibinfo{pages}{3}.
\newblock


\bibitem[Zehlike et~al\mbox{.}(2022)]%
        {zehlike2022fair}
\bibfield{author}{\bibinfo{person}{Meike Zehlike}, \bibinfo{person}{Tom
  S{\"u}hr}, \bibinfo{person}{Ricardo Baeza-Yates}, \bibinfo{person}{Francesco
  Bonchi}, \bibinfo{person}{Carlos Castillo}, {and} \bibinfo{person}{Sara
  Hajian}.} \bibinfo{year}{2022}\natexlab{}.
\newblock \showarticletitle{Fair Top-k Ranking with multiple protected groups}.
\newblock \bibinfo{journal}{\emph{Information Processing \& Management}}
  \bibinfo{volume}{59}, \bibinfo{number}{1} (\bibinfo{year}{2022}),
  \bibinfo{pages}{102707}.
\newblock


\bibitem[Zhang et~al\mbox{.}(2014)]%
        {optimalbiding}
\bibfield{author}{\bibinfo{person}{Weinan Zhang}, \bibinfo{person}{Shuai Yuan},
  {and} \bibinfo{person}{Jun Wang}.} \bibinfo{year}{2014}\natexlab{}.
\newblock \showarticletitle{Optimal real-time bidding for display advertising}.
  In \bibinfo{booktitle}{\emph{Proceedings of the 20th ACM SIGKDD international
  conference on Knowledge discovery and data mining}}. ACM,
  \bibinfo{pages}{1077--1086}.
\newblock


\bibitem[Zwicker(2016)]%
        {DBLP:reference/choice/Zwicker16}
\bibfield{author}{\bibinfo{person}{William~S. Zwicker}.}
  \bibinfo{year}{2016}\natexlab{}.
\newblock \showarticletitle{Introduction to the Theory of Voting}.
\newblock In \bibinfo{booktitle}{\emph{Handbook of Computational Social
  Choice}}, \bibfield{editor}{\bibinfo{person}{Felix Brandt},
  \bibinfo{person}{Vincent Conitzer}, \bibinfo{person}{Ulle Endriss},
  \bibinfo{person}{J{\'{e}}r{\^{o}}me Lang}, {and} \bibinfo{person}{Ariel~D.
  Procaccia}} (Eds.). \bibinfo{publisher}{Cambridge University Press},
  \bibinfo{pages}{23--56}.
\newblock
\urldef\tempurl%
\url{https://doi.org/10.1017/CBO9781107446984.003}
\showDOI{\tempurl}


\end{thebibliography}

\end{document}